# A possible mechanism for high temperature superconductivity in the cuprates

Stanley Engelsberg
Physics Department
University of Massachusetts
Amherst, Massachusetts, 01003

Several ideas that have been shown to apply to superconductors and the cuprates in particular are joined together to form a mechanism for high temperature superconductivity. The mechanism is basically a weak BCS (1) type coupling between the holes formed in doping the crystal and the optical modes that can be excited when the stripes that are formed beat against one another. The fundamental energy scale set by the normal modes of the oscillating stripes (strions) is of the order of an electron volt. Thus the mechanism allows for room temperature superconductivity. Several important experimental observations are in agreement with the predictions of this theory. Direct observation of the strions might be possible in electron energy loss experiments.

A great deal of energy has been expended in trying to understand the superconductivity found in a number of the cuprates at temperatures above that of liquid nitrogen. Not altogether surprising given the many technological advances that would follow from room temperature superconductivity. This work makes use of a number of well accepted ideas to suggest a possible mechanism for the observed high temperature superconductivity in the cuprates. The fundamental idea is to couple the holes that are formed in the fabrication of the compounds to the normal modes of the stripes that the holes join together to form.

The existence of stripes was suggested using theoretical arguments and confirmed experimentally in some of the cuprates and nickelates having a perovskite structure. The article by Emery, Kivelson and Tranquada (2) gives a clear and thought provoking summary of the status of stripe phases in high temperature superconductors as of 1999 and includes a number of important references to the theory and experiments mentioned above.

The starting point will be to estimate the normal mode frequency associated with the stripes beating against one another. One might expect this frequency to determine the characteristic energy of the coupling of the holes to the stripe oscillations (strions). There is a direct analogy to the three dimensional oscillations of an electron plasma that produces the plasmon frequency,

$$\omega^2 \text{ (plasmon)} = n_3 \cdot e^2/m.$$

$n_3$ is the number density (electrons per unit volume) in the plasma. The energy loss associated with electrons emitting a plasmon in a metal is typically 10 eV. The characteristic strion mode frequency is given by

$$\omega^2 \text{ (strion)} = n_1 \cdot e^2 / (\pi \, m \, L^2)$$

$n_1$ is the one dimensional number density (holes per unit length) in the stripe and L is the distance between stripes. Since the number of holes is less than one per atom in the copper oxide planes and L is greater than the interatomic spacing, one might expect the characteristic strion energy to be somewhat less than the plasmon energy, say of the order of 1 eV. The superconducting transition temperature of the high temperature superconductors (~100K) corresponds to energy of about 8 mev. Thus, we would expect weak coupling BCS theory (1) to be appropriate

$$k_B T_c \approx \hbar \omega \text{(strion)} \exp(-1/\lambda).$$

The parameter $\lambda$ is dimensionless and represents the strength of the hole-strion interaction. In the weak coupling limit BCS theory (1) predicts and tunneling experiments show no details of the interaction coupling, e.g. the phonon structure that leads to electron pairing. For the typical values of the parameters given above, $\lambda$ is about 0.2, a number that is typical of weak coupling metallic superconductors, e.g. Al and Zn.

There are a number of obvious predictions of this theory. Perhaps the most striking is that there is no isotope effect. In its place there is a linear relation between the superconducting transition temperature and the inverse of the distance between the stripes, $1/L$. Both of these implications are experimentally observed for $YBa_2Cu_3O_{6+x}$ (3, 4). Both a $1/L$ dependence (5) and an isotope effect (6, 7) are observed for the transition temperature of $La_{2-x}Sr_xCuO_4$. The isotope effect arises through a simple and natural strion-phonon coupling as shown below.

Since the crystal lattice is seen to be modulated by the appearance of stripes (2, 4, 5, 8), we can expect the normal modes of the stripes to be coupled to the normal modes of the lattice, the phonons. Raveau, et.al. (8) present a very large number of high resolution electron microscope and x-ray diffraction pictures of the modulations (stripes) that appear in all of the superconductors that they examine. We take the strion-phonon coupling into account most simply by putting a coupling term, g, in the off diagonal matrix elements of the 2 by 2 Hamiltonian matrix whose unperturbed eigenvalues are the characteristic strion frequency, $\omega_S$, and the characteristic phonon frequency, $\omega_D$. Diagonalizing this Hamiltonian gives two solutions,

$$\omega = (\omega_S + \omega_D)/2 \pm \sqrt{\{(\omega_S - \omega_D)^2/4 + g^2\}}.$$

When the ion displacement coupling to the stripes is very weak, g approaches zero, and the normal modes are the original strion and phonon frequencies. In such cases the strion frequency, being much higher than the phonon frequency, will appear in the BCS relation for Tc, as before. For crystals in which the coupling between the two motions becomes stronger the characteristic frequency determining the transition temperature will be the larger of the two eigenvalues

$$\omega = (\omega_S + \omega_D)/2 + \sqrt{\{(\omega_S - \omega_D)^2/4 + g^2\}}.$$

We see that the interaction of strion and phonon modes leads to a superconducting transition temperature that has a dependence on both the distance between stripes and the isotopic mass of the ions (5,6,7), but not the simple $1/L$ or $1/\sqrt{M}$ dependence seen in the pure strion or pure phonon weak coupling limit.

The dependence of the strion frequency on $\sqrt{n_l}$ as reflected in the dependence of the transition temperature on hole doping is a more delicate question because of our incomplete knowledge about the stripe stability as a function of doping. It would be interesting to observe the excitation of strions with electron energy loss experiments as is done for plasmon excitations.  The electron energy loss experiments of Nücker et. al. (9) on the lanthanum strontium copper oxide and yttrium barium copper oxide compounds show peaks and shoulders that shift with hole concentration in the direction predicted by our theory.  The energy scale between various features in the energy loss spectra is of order 1 eV, which is about the value we estimated previously for the excitation of a strion.

The glaring weakness of our approach is the implicit relation between $1/L$ and $\sqrt{n_1} \exp(-1/\lambda)$ that might appear in a completely self consistent theory of holes, stripes and superconductivity in layered antiferromagnetic transition metal oxides.  Perhaps this weakness is no greater than that in the application of BCS weak coupling theory (1) to ordinary superconducting metals.  If our theory is correct there is a renewed possibility for a room temperature superconductor in which the hole-strion interaction is somewhat stronger than the weak coupling limit that has been discovered so far.

I would like to thank R. Cava, R. Guyer, J. Machta and N. Prokof'ev for helpful discussions.